\documentclass[conference]{IEEEtran}
\usepackage[left=0.67in,right=0.67in,top=0.64in,bottom=0.70in]{geometry}
\usepackage[T1]{fontenc}
\usepackage[latin9]{inputenc}
\usepackage{amsthm}
\usepackage{amsmath}
\usepackage{graphicx} 
\usepackage{color} 
\usepackage{cite}
\usepackage{algpseudocode}
\usepackage{algorithm}
\usepackage{amssymb}
\usepackage{subfigure}
\usepackage{esint}
\usepackage{algpseudocode}
\usepackage{epstopdf}
\usepackage{multirow}
\usepackage{makecell}
\usepackage{float}
\usepackage{lipsum}
\usepackage{balance}
\usepackage{ctable}
\usepackage{nicefrac}

\theoremstyle{plain}
\theoremstyle{plain}

\newcommand{\comment}[1]{}

\IEEEoverridecommandlockouts

\allowdisplaybreaks

\begin{document}

\title{LDPC Decoders Prefer More Reliable Parity Bits: Unequal Data Protection Over BSC \vspace{-0.4em}
}








\author{
   \IEEEauthorblockN{Beyza Dabak\IEEEauthorrefmark{1}, Ece Tiryaki\IEEEauthorrefmark{2}, Robert Calderbank\IEEEauthorrefmark{1}, and Ahmed Hareedy\IEEEauthorrefmark{2}}
   \IEEEauthorblockA{\IEEEauthorrefmark{1}Electrical and Computer Engineering Department, Duke University, Durham, NC 27708 USA \\ \IEEEauthorrefmark{2}Electrical and Electronics Engineering Department, Middle East Technical University, Ankara, 06810 Turkey \\ beyza.dabak@duke.edu, tiryaki.ece@metu.edu.tr, robert.calderbank@duke.edu, and ahareedy@metu.edu.tr\vspace{-1.5em}
   }
}
\maketitle

\begin{abstract}
Low-density parity-check (LDPC) codes are specified by graphs, and are the error correction technique of choice in many communications and data storage contexts. Message passing decoders diffuse information carried by parity bits into the payload, and this paper measures the value of engineering parity bits to be more reliable than message bits. We consider the binary symmetric channel (BSC) and measure the impact of unequal data protection on the threshold of a regular LDPC code. Our analysis also includes doping where the parity bits are known to the decoder. We investigate BSC with Gallager-A decoder, with a $3$-level-alphabet decoder, and with a full~belief propagation decoder. We demonstrate through theoretical analysis and simulation that non-equiprobable inputs lead to significant improvements both in the threshold and in the speed with which the decoder converges. We also show that all these improvements are possible even with a simple $3$-level-alphabet decoder.
\end{abstract}





\section{Introduction}\label{sec_intro}

Since their introduction by Gallager \cite{gallager} in 1962, low-density parity-check (LDPC) codes have found widespread use in wireless communication and data storage. It is worthwhile to consider how to improve decoder performance by directly shaping the reliability of inputs. Here, for unequal data protection (UDP), a specific part of the transmitted/stored data has higher reliability by design. Moreover, doping is to have perfect-reliability nodes, i.e., nodes corresponding to bits that are known to the decoder, in the LDPC codeword. In UDP, one prior approach is to protect message bits rather than parity bits \cite{furzun_uep,zhu_doping}, but a high rate code will have many more message bits to protect. We choose to protect parity bits rather than message bits of a regular LDPC code in order to have the decoder experience a better channel. A different approach entirely is to engineer unequal reliabilities by shaping the LDPC code, for example, through the choice of degree distribution in multi-edge type (MET) LDPC codes \cite{met}, or through the optimization of code irregularities (node degrees) in order to obtain several classes of protection within the codeword \cite{fekri,sara,declerq}, or through doping to improve performance via creating low-degree check nodes (CNs) \cite{sok_doping}. Our approach, instead, is independent of the specific LDPC code construction.

The value of protecting parity bits over message bits was demonstrated empirically in the context of data storage \cite{ahh_loco}. In magnetic recording (MR), a lexicographically-ordered constrained (LOCO) code \cite{ahh_general} served as an inner code that protected the parity bits of an outer LDPC code by mitigating inter-symbol interference. Note that the inner code decreases the overall code rate. Density gains of up to $16\%$ were observed with respect to an LDPC coded system with no inner code and with the same overall rate. The empirical results presented in \cite{ahh_loco} are the motivation behind this paper as we want to demonstrate the threshold gains of this UDP setup theoretically. The \textit{threshold} is a measure of the highest channel noise/error level below which, coded error-free transmission is possible given a specific set of code parameters and a decoder \cite{de}. Other than in MR, UDP and doping can also be applied in Flash memories to preserve access speed and in wireless communications to increase the transmission rate for the same performance \cite{ahh_qaloco}.

In our theoretical analysis of LDPC decoding, we operate on the binary symmetric channel (BSC) and subject parity bits to lower crossover probabilities than message bits. We employ density evolution (DE) \cite{de}, an analytical tool widely used to evaluate asymptotic performance and investigate convergence behavior of iterative decoders. DE tracks the probability mass/density of messages passed in the code graph from iteration to iteration. In \cite{exact}, exact thresholds for Gallager A algorithm are determined for BSC. The methods introduced in \cite{de} and \cite{exact} have been widely used in the literature to analytically obtain thresholds for various codes, channels, and decoders \cite{varshney1, zarrin, paolini, vatta}.

We analyze the BSC with finite alphabet decoders (the $2$-level Gallager-A decoder and the $3$-level error-and-erasures decoder), also with an infinite alphabet decoder (the full belief propagation (BP) decoder). UDP and doping provide significant performance gains, even with a simple quantized decoder. Performance gains jump as we increase the decoder alphabet from $2$ to $3$ levels, and in $3$-level decoder UDP is approximately as effective as it is in full BP. With UDP, we achieve up to $38.2\%$ and $42.7\%$ threshold gains for $3$-level and BP decoders, respectively.

Our contributions are as follows. First, we derive the DE equations for UDP and doping setups for all decoders (adopting the analysis in \cite{de}). Second, we further analyze the iterative algorithm performance to provide insight into decoder dynamics. In certain cases, we are able to derive closed-forms of the exact threshold, tight closed-form approximations of the threshold in terms of code parameters, or probability approximations that depend on the iteration index. Third, we use the aforementioned DE equations to numerically find the exact threshold for different decoders. Fourth, we experimentally demonstrate the gains of the UDP and doping setups. We also show that, at the same noise level, unequal data protection improves the speed with which the decoder converges.

The rest of the paper is organized as follows. In Section~\ref{sec_prelim}, we discuss the preliminaries. Moving from hard to soft-decision decoding, each decoding algorithm is discussed in a separate section that is self-contained with its respective DE analysis, analytical calculations, and experimental results. Section~\ref{sec_conc} concludes the paper. 

\section{Preliminaries}\label{sec_prelim}

We introduce our UDP setup, the decoders, and the uniform (no UDP case) DE equations of the decoders available in the literature \cite{de}.

\subsection{UDP Setup}
We work on regular $(d_{\textup{v}},d_{\textup{c}})$ LDPC codes, where $d_{\textup{v}}$ and $d_{\textup{c}}$ are the variable node (VN) degree and the CN degree, respectively. In UDP setup, parity bits are more reliable than the message bits, i.e., parity and message bits effectively see two different channels with different reliabilities. Let $x$ be the fixed number of VNs, connected to a specific CN, that are reliable. Thus per CN, there are $x$ reliable VNs and  $d_{\textup{c}}-x$ regular VNs. Since the fraction of parity VNs is $d_{\textup{v}}/d_{\textup{c}}$ and the fraction of reliable VNs is $x/d_{\textup{c}}$, it is natural to assume that $x = d_{\textup{v}}$ without loss of generality. We denote the error probabilities of parity bits and message bits by $\overline{p}$ and $p$, respectively. Note that we do not change the code construction in our UDP setup contrarily to other works \cite{met,fekri,sara,declerq}. Here, each VN representing a message bit is connected to $x$ VNs representing parity bits through each of its CNs.~This~ensures the diffusion of higher fidelity messages from parity to message bits. We compare the performance of UDP setup against the uniform setup where all bits have the same error probability, $p$. The data symbols are binary in $\{0, 1\}$, where $0$ ($1$) is mapped to $1$ ($-1$). Without loss of generality since the system is symmetric, we assume all-zero codewords are transmitted.

\subsection{BSC - Gallager A Decoder}
For a BSC with message alphabet $k  \in \{-1,1\}$, let $p_{k}^{(l)}$ denote the probability that the message sent from a VN to a CN at iteration $l$ equals $k$. In Gallager A decoder, a VN value is flipped if and only if all CNs that are connected to it agree it should be flipped at an iteration. As illustrated by Gallager \cite{gallager}, the evolution of $p_{-1}^{(l)}$ under Gallager A decoder, which is the probability of error here, is given by: 
\begin{align} \label{gallager_a}
p_{-1}^{(l)} &= p_{-1}^{(0)} - p_{-1}^{(0)}A^{d_{\textup{v}}-1}  + (1-p_{-1}^{(0)})B^{d_{\textup{v}}-1}, 
\end{align}
where
\begin{align}
A = \bigg[\frac{1+(1-2p_{-1}^{(l-1)})^{d_{\textup{c}}-1}}{2} \bigg], \textup{ } B = \bigg[\frac{1-(1-2p_{-1}^{(l-1)})^{d_{\textup{c}}-1}}{2} \bigg]. \nonumber
\end{align}

\subsection{BSC - $3$-Level-Alphabet Decoder}
The $3$-level decoder introduces discretized soft-decision into the algorithm. The alphabet is extended into $k \in \{-1,0,1\}$. The message at the VN is the sign of the real sum of the received message times an appropriate weight $w^{(l)}$ (for the $l^{\textup{th}}$ iteration) and the messages coming from $d_{\textup{v}}-1$ CNs. Whereas the message at the CNs is the product of messages coming from $d_{\textup{c}}-1$ VNs. More details can be found in \cite{mitz}. 

Let $p_{k}^{(l)}$ ($q_{k}^{(l)}$) denote the probability that the message sent from a VN to a CN (from a CN to a VN) at iteration $l$ equals $k$. The DE equations \cite{de}:
\begin{align}\label{eqn_q_bsc_unif}
q_{0}^{(l)} &= 1-(1 - p_{0}^{(l-1)})^{dc-1} \nonumber \\
q_{1}^{(l)} &= \frac{1}{2} \bigr[(p_{1}^{(l-1)} + p_{-1}^{(l-1)})^{dc-1} + (p_{1}^{(l-1)} - p_{-1}^{(l-1)})^{dc-1}\bigr] \nonumber \\
q_{-1}^{(l)} &= \frac{1}{2} \bigr[(p_{1}^{(l-1)} + p_{-1}^{(l-1)})^{dc-1} - (p_{1}^{(l-1)} - p_{-1}^{(l-1)})^{dc-1}\bigr].
\end{align}
\begin{align} \label{eqn_p_bsc_unif}
p_{k'}^{(l)} &= p_{0}^{(0)} \sum_{(i,j):i-j \bigtriangleup 0}\binom{d_{\textup{v}}-1}{i,j,\zeta}\big(q_{1}^{(l)}\big)^{i}\big(q_{-1}^{(l)}\big)^{j}\big(q_{0}^{(l)}\big)^{\zeta}&&\nonumber \\
&\hspace{+1.0em} + p_{1}^{(0)} \sum_{(i,j):i-j \bigtriangleup -w_{l}}\binom{d_{\textup{v}}-1}{i,j,\zeta} \big(q_{1}^{(l)}\big)^{i}\big(q_{-1}^{(l)}\big)^{j}\big(q_{0}^{(l)}\big)^{\zeta}&&\nonumber \\
&\hspace{+1.0em} + p_{-1}^{(0)} \sum_{(i,j):i-j \bigtriangleup w_{l}}\binom{d_{\textup{v}}-1}{i,j,\zeta} \big(q_{1}^{(l)}\big)^{i}\big(q_{-1}^{(l)}\big)^{j}\big(q_{0}^{(l)}\big)^{\zeta} \nonumber \\
p_{-1}^{(l)} &= 1-p_{0}^{(l)}-p_{1}^{(l)},
\end{align}
where $k' \in \{0,1\}$, $\zeta=d_{\textup{v}}-1-i-j$, and the operator $\bigtriangleup$ is equality ($=$) if $k'=0$, and is greater than ($>$) if $k'=1$.

\subsection{BSC - BP Decoder} \label{de_bp_uniform}
We now discuss DE for the case when message alphabet is continuous \cite{de}. In BP, the messages sent on an edge roughly represent posterior densities on the incident variable node. The conditional probability density associated with an edge is defined by the probability pair ($p_{-1}, p_{1})$, where $p_{-1}+p_{1}=1$, and it can be represented by corresponding log-likelihood ratio (LLR) $\log\frac{p_1}{p_{-1}}$. Transferred messages between nodes are LLRs, which are real random variables, and they are assumed to be independent. The message sent from a VN to a CN is the sum of the channel message and $(d_{\textup{v}}-1)$ messages coming from other adjacent CNs, and its probability density is calculated by the convolution ($P_{0} \otimes P_{1} \otimes \dots \otimes P_{d_{\textup{v}}-1}$), where $P_{i}$ denotes the probability density of message $i$.

When the message sent from a CN to a VN is considered, the message going to the CN from edge $t \in [d_{\textup{c}}-1]$ is the LLR $\log\frac{p_{1}^{t}}{p_{-1}^{t}} = \log\frac{p_{0}^{t}}{p_{1}^{t}}$.\footnote{We follow the notation in \cite{de} here. Recall that the binary field $\{0,1\}$ corresponds $\{1,-1\}$.} The outgoing message from a CN along an edge is the LLR $\log\frac{p_{0}'}{p_{1}'}$. $p_{0}'$ ($p_{1}'$) denotes the probability that mod 2 sum of  $d_{\textup{c}}-1$ independent random variables is $0$ ($1$), and can be calculated using the cyclic convolution of these random variables \cite{de}. The update law is specified as follows:
\begin{align} \label{bp_de_cn}
p_{0}' - p_{1}'&= \hspace{-0.2em}\prod_{t=1}^{d_{\textup{c}}-1} \big(p_{0}^{t} - p_{1}^{t} \big).
\end{align}

In order to study the evolution of densities at the CN side, the probability density $(p_{0},p_{1})$ can be represented in $\textup{GF}(2) \times [0,\infty)$ as follows:
\begin{align} \label{representation}
(\textup{lgsgn}(p_{0}-p_{1}), -\log|p_{0}-p_{1}|).
\end{align}
The $\textup{lgsgn}$ is defined as 1 (0) if the difference is less (more) than $0$. For $L=\log\frac{p_{0}}{p_{1}}$, $p_{0}-p_{1} = \frac{e^{L}-1}{e^{L}+1}$ can be substituted in (\ref{representation}). With this representation, the CN-to-VN message can be computed via addition, which means the density can be again computed via convolution. By taking fast Fourier transform (FFT) of the magnitude term densities, multiplying transforms, applying inverse FFT (IFFT), then changing the measure backward, the CN-to-VN message density can be computed.

\section{BSC - Gallager A Decoder}\label{sec_bsc_b}
We investigate UDP and doping gains on BSC with Gallager A decoder. We derive the DE equations for UDP setup, exact doping threshold using the results in \cite{exact} for cases, tight closed-form uniform and doping threshold approximations as a function of $d_{\textup{v}}$ and $d_{\textup{c}}$ for other cases, and numerically demonstrate doping threshold gains.

\subsection*{DE Equations for UDP}
Let $\overline{p}_{k}^{(l)}$ ($p_{k}^{(l)}$), for $k  \in \{-1,1\}$, denote the probability that the message sent at iteration $l$ equals $k$ for reliable (regular) VNs. Let $x = d_{\textup{v}}$ be the number of reliable VNs connected to a specific CN. The (iterative) evolution of $\overline{p}_{-1}^{(l)}$ and $p_{-1}^{(l)}$ under Gallager A decoder is as follows:
\begin{align} \label{gallager_a_udp}
\widetilde{p}_{-1}^{(l)} &= \widetilde{p}_{-1}^{(0)} - \widetilde{p}_{-1}^{(0)}\bigg[\frac{1+(1-2\overline{p}_{-1}^{(l-1)})^{\alpha}(1-2p_{-1}^{(l-1)})^{\beta}}{2} \bigg]^{d_{\textup{v}}-1} \nonumber && \\
&\hspace{+1.0em} + (1-\widetilde{p}_{-1}^{(0)})\bigg[\frac{1-(1-2\overline{p}_{-1}^{(l-1)})^{\alpha}(1-2p_{-1}^{(l-1)})^{\beta}}{2} \bigg]^{d_{\textup{v}}-1} \hspace{-1em}, 
\end{align}
where 
\begin{align}
(\alpha, \beta) = \left\{\begin{matrix}(x,d_{\textup{c}}-1-x), \textup{ } &\widetilde{p}_{-1}^{(l)} = p_{-1}^{(l)},
\\ (x-1,d_{\textup{c}}-x), \textup{ } &\widetilde{p}_{-1}^{(l)} = \overline{p}_{-1}^{(l)}.
\end{matrix}\right.
\end{align}

As an extreme case of UDP, we consider doping when the parity bits are known to the decoder, i.e., $\overline{p}_{-1}^{(l)} = 0$, $\overline{p}_{1}^{(l)} = 1$, for all iterations, and
\begin{align} \label{doping_de}
{p}_{-1}^{(l)} &= {p}_{-1}^{(0)} - {p}_{-1}^{(0)}\bigg[\frac{1+(1-2p_{-1}^{(l-1)})^{d_{\textup{c}}-1-x}}{2} \bigg]^{d_{\textup{v}}-1} \nonumber && \\
&\hspace{+1.0em} + (1-{p}_{-1}^{(0)})\bigg[\frac{1-(1-2p_{-1}^{(l-1)})^{d_{\textup{c}}-1-x}}{2} \bigg]^{d_{\textup{v}}-1}.
\end{align}

\subsection*{Exact Thresholds for Doping}

Exact thresholds for BSC with Gallager A decoding are derived for the uniform case in \cite{exact}. If we consider $p_{-1}^{(l)}$ as a function of $p_{-1}^{(l-1)}$, it is a decreasing function for any converging case ($p_{-1}^{(l)} < p_{-1}^{(l-1)}$). In \cite{exact}, the authors prove that the uniform threshold is the minimum of the variable $\gamma$ given in equation (\ref{bazzi_eqn}) and the smallest root $\eta$ of the exact polynomial in (\ref{poly_unf}).

The variable $\gamma$ is derived from the equation describing the fixed point at the end of the decoding process, i.e., the point at which $p_{-1}^{(l)} = p_{-1}^{(l-1)}$. This analysis leads to: 
\begin{align} \label{bazzi_eqn}
    \gamma &=\frac{1-\lambda_{2}\rho'(1)}{\lambda'(1)\rho'(1)-\lambda_{2}\rho'(1)},
\end{align}
where $\lambda(y) := \sum_{i=2}^{\infty}\lambda_{i}y^{i-1}$ and $\rho(y):=\sum_{i=2}^{\infty}\rho_{i}y^{i-1}$ are the VN and CN degree distribution functions, respectively. 

For a regular $(d_{\textup{v}},d_{\textup{c}})$ LDPC code, $\lambda(y) = y^{d_{\textup{v}}-1}$ while $\rho(y) = y^{d_{\textup{c}}-1}$. For the case of interest where $d_{\textup{v}} \geq 3$, $\lambda_2 = 0$. Therefore, (\ref{bazzi_eqn}) becomes:
\begin{align}\label{bazzi_reg}
    \gamma &=\frac{1}{\lambda'(1)\rho'(1)} = \frac{1}{(d_{\textup{v}}-1)(d_{\textup{c}}-1)}.
\end{align}

The variable $\eta$ is obtained from beginning of the decoding process, specifically where the curve $p_{-1}^{(l)}=f(p_{-1}^{(l-1)})$ intersects with the $45$-degree line. Therefore, we can write $p_{-1}^{(1)}$ as a function of $p_{-1}^{(0)}$ in (\ref{gallager_a}), then set $p_{-1}^{(1)} = p_{-1}^{(0)}$ to reach the polynomial below, which we seek its smallest root $\eta$:
\begin{align} \label{poly_unf}
p(y) \hspace{-0.2em}&:= \hspace{-0.1em}y\bigg[\frac{1\hspace{-0.2em}+\hspace{-0.2em}(1\hspace{-0.2em}-\hspace{-0.2em}2y)^{k}}{2} \bigg]^{d_{\textup{v}}-1} \hspace{-1.1em}+ \hspace{-0.1em}(y\hspace{-0.2em}-\hspace{-0.2em}1)\bigg[\frac{1\hspace{-0.2em}-\hspace{-0.2em}(1\hspace{-0.2em}-\hspace{-0.2em}2y)^{k}}{2} \bigg]^{d_{\textup{v}}-1} \hspace{-1em},
\end{align}
where $k= d_{\textup{c}}-1$ and our focus in this work is on regular LDPC codes. 

Now, we discuss UDP and doping. Observe that the term $\lambda'(1)\rho'(1)$ in (\ref{bazzi_eqn}) and (\ref{bazzi_reg}) stems from the derivative of
\begin{align}\label{pplus_eqn}
p^+(y) = \lambda \left[ \frac{1+\rho(1-2y)}{2} \right ]
\end{align}
at $y=0$ (see \cite{exact}), which can be seen using (\ref{gallager_a}). For regular nodes in the UDP setup, we define $\rho_1(y) = y^{d_{\textup{c}}-1-x}$ and $\rho_2(y)=y^{x}$, while $\lambda(y) =y^{d_{\textup{v}}-1}$ stays the same. Using (\ref{gallager_a_udp}) and (\ref{doping_de}), we rewrite (\ref{pplus_eqn}) for the doping case as follows:
\begin{align}\label{pplus_dope}
p^+(y) = \lambda \left[ \frac{1+\rho_1(1-2y)\rho_2(1)}{2} \right ] = \lambda \left[ \frac{1+\rho_1(1-2y)}{2} \right ].
\end{align}
Now, the derivative of $p^+(y)$ at $y=0$ results in $\lambda'(1)\rho'_1(1)$. Therefore, and in a way similar to what is done to reach (\ref{bazzi_reg}), we get the updated $\gamma$ for regular nodes:
\begin{align}\label{ours_reg}
    \gamma &=\frac{1}{\lambda'(1)\rho'_1(1)} = \frac{1}{(d_{\textup{v}}-1)(d_{\textup{c}}-1-x)}.
\end{align}
Since we are seeking the average $\gamma$, $\overline{\gamma}$, on all nodes (regular and reliable/known), we reach:
\begin{align} \label{th_doping}
    \overline{\gamma} &= \frac{1}{(d_{\textup{v}}-1)(d_{\textup{c}}-1-x)}\biggr(1-\frac{d_{\textup{v}}}{d_{\textup{c}}}\biggr).
\end{align}

The \textit{exact threshold} of the doping setup is then the minimum of $\overline{\gamma}$ and the smallest root to the exact doping polynomial (weighted in a manner similar to that of $\overline{\gamma}$, and denote by $\overline{\eta}$). Note that, the exact doping polynomial is the same as that in (\ref{poly_unf}) except for that $k= d_{\textup{c}}-1-x$ from equation (\ref{doping_de}).

We observe in Table~\ref{table_gall} that for certain values of $d_{\textup{v}}$, $\overline{\gamma}$ from (\ref{th_doping}) is the exact threshold. For example, when $d_{\textup{v}}=5, d_{\textup{c}}=10$, $\overline{\eta}=0.0830$ is computed from the polynomial, while $\overline{\gamma}=0.0313$ and the threshold, obtained numerically, is $p_{\textup{Doping}} = \min\{\overline{\gamma},\overline{\eta}\} =0.0313$. 

\begin{table*}
\caption{Thresholds, Approximations, and Threshold Gains of Doping for Various $(d_{\textup{v}},d_{\textup{c}})$ LDPC Codes \\ On BSC with Gallager A Decoder}
\vspace{-0.5em}
\centering
\begin{tabular}{|c|c|c!{\vrule width 0.1em}c|c|c|c!{\vrule width 0.1em}c|c|c|c!{\vrule width 0.1em}c|}
\hline
\multicolumn{3}{|c!{\vrule width 0.1em}}{\makecell{}} & \multicolumn{4}{c!{\vrule width 0.1em}}{\makecell{\textbf{Uniform}}} & \multicolumn{4}{c!{\vrule width 0.1em}}{\makecell{\textbf{Doping}}} & \\
\hline
$d_{\textup{v}}$ & $d_{\textup{c}}$ & Rate & $p_{\textup{UNF}}$ & $\gamma$ & $\eta$ & $p_{\textup{UNF}}^{\textup{a}}$ & $p_{\textup{Doping}}$ & $\overline{\gamma}$ & $\overline{\eta}$ & $p_{\textup{Doping}}^{\textup{a}}$ & \textbf{Gain}   \\
\specialrule{.1em}{.05em}{.05em} 
$3$ & $10$ & $0.70$ & $0.0123$ & $0.0556$ & $0.0123$ & $0.0104$ & $0.0193$ & $0.0583$ & $0.0193$ & $0.0155$ & $56.9\%$ \\
\hline
$3$ & $15$ & $0.80$ & $0.0051$ & $0.0357$ & $0.0051$ & $0.0045$ & $0.0066$ & $0.0364$ & $0.0066$ & $0.0057$ & $29.4\%$ \\
\specialrule{.1em}{.05em}{.05em} 
$4$ & $15$ & $0.73$ & $0.0194$ & $0.0238$ & $0.0194$ & $0.0151$ & $0.0237$ & $0.0244$ & $0.0237$ & $0.0181$ & $22.2\%$ \\
\hline 
$4$ & $30$ & $0.87$ & $0.0065$ & $0.0115$ & $0.0065$ & $0.0053$ &  $0.0070$ & $0.0116$ & $0.0070$ & $0.0057$ & $7.7\%$ \\
\specialrule{.1em}{.05em}{.05em} 
$5$ & $10$ & $0.50$ & $0.0278$ & $0.0278$ & $0.0569$ & -- & $0.0313$ & $0.0313$ & $0.0830$ & -- & $12.6\%$ \\
\hline
$5$ & $15$ & $0.67$ & $0.0179$ & $0.0179$ & $0.0313$ & -- & $0.0185$ & $0.0185$ & $0.0379$ & -- & $3.4\%$ \\
\hline
\end{tabular}
\label{table_gall}
\end{table*}

\begin{figure*}
\begin{align} \label{gall_a_unf_approx}
p_{-1}^{(l)} &\approx p_{-1}^{(0)}\Big(\hspace{-0.1em}1\hspace{-0.1em}-\hspace{-0.1em}\big(1\hspace{-0.1em}-\hspace{-0.1em}(d_{\textup{c}}\hspace{-0.1em}-\hspace{-0.1em}1)p_{-1}^{(l-1)} \hspace{-0.1em}+\hspace{-0.1em} (d_{\textup{c}}\hspace{-0.1em}-\hspace{-0.1em}1)(d_{\textup{c}}\hspace{-0.1em}-\hspace{-0.1em}2)(p_{-1}^{(l-1)})^{2}\big)^{d_{\textup{v}}-1}\hspace{-0.1em}\Big) \hspace{-0.2em}+\hspace{-0.2em}
(1\hspace{-0.1em}-\hspace{-0.1em}p_{-1}^{(0)})\Big(\hspace{-0.1em}(d_{\textup{c}}\hspace{-0.1em}-\hspace{-0.1em}1)p_{-1}^{(l-1)}\hspace{-0.1em} + \hspace{-0.1em}(d_{\textup{c}}\hspace{-0.1em}-\hspace{-0.1em}1)(d_{\textup{c}}\hspace{-0.1em}-\hspace{-0.1em}2)(p_{-1}^{(l-1)})^{2}\hspace{-0.1em}\Big)^{d_{\textup{v}}-1} \nonumber \\
&\approx p_{-1}^{(0)}\Big(1-1+(d_{\textup{v}}-1)(d_{\textup{c}}-1)p_{-1}^{(l-1)}-(d_{\textup{v}}-1)(d_{\textup{c}}-1)(d_{\textup{c}}-2)(p_{-1}^{(l-1)})^{2}-\frac{1}{2}(d_{\textup{v}}-1)(d_{\textup{v}}-2)(d_{\textup{c}}-1)^{2}(p_{-1}^{(l-1)})^{2}\Big) \nonumber && \\
&\hspace{+1.0em}- p_{-1}^{(0)}(d_{\textup{c}}-1)^{d_{\textup{v}}-1}(p_{-1}^{(l-1)})^{d_{\textup{v}}-1} + (d_{\textup{c}}-1)^{d_{\textup{v}}-1}(p_{-1}^{(l-1)})^{d_{\textup{v}}-1}.
\end{align}
\rule{18cm}{0.4pt}
\end{figure*}

\subsection*{Tight Approximations}
\subsubsection{Uniform Setup}
Bazzi \emph{et. al.} find the exact smallest root $\eta$ of the polynomial in (\ref{poly_unf}) numerically \cite{exact}. To derive a closed-form approximate equation of the threshold, in (\ref{gall_a_unf_approx}), we approximate (\ref{poly_unf}) that is based on (\ref{gallager_a}) by taking only $3$ binomial expansion terms to represent $(1-2y)^k$, where $k={d_{\textup{c}}-1}$, set  $y=p_{-1}^{(0)}$, and then solve the quadratic equation resulting from setting the approximate polynomial to $0$. The smallest solution of (\ref{gal_closed}) with $k = d_{\textup{c}}\hspace{-0.1em}-\hspace{-0.1em}1$ gives an approximate threshold value in specific cases (where this solution is less than ${\gamma}$).
\begin{align} \label{gal_closed}
0 &= (d_{\textup{v}}-1)kp_{-1}^{(0)} - (d_{\textup{v}}-1)k\bigg(\frac{kd_{\textup{v}}}{2}-1\bigg)(p_{-1}^{(0)})^{2} && \nonumber \\
&\hspace{+1.0em}- k^{d_{\textup{v}}-1}(p_{-1}^{(0)})^{d_{\textup{v}}-1} + k^{d_{\textup{v}}-1}(p_{-1}^{(0)})^{d_{\textup{v}}-2} -1.
\end{align}

Consider the case of $d_{\textup{v}}=3$ (in which, Gallager A becomes equivalent to Gallager B) and plug $k = d_{\textup{c}}-1$ in (\ref{gal_closed}), 
\begin{align}
0 &= (2d_{\textup{c}}-2)(2d_{\textup{c}}-3)(p_{-1}^{(0)})^{2} - (d_{\textup{c}}^{2}-1)p_{-1}^{(0)} +1, \nonumber
\end{align}
\begin{align}
p_{-1}^{(0)} &=  \frac{d_{\textup{c}}^{2}-1 \pm \sqrt{d_{\textup{c}}^{4}-18d_{\textup{c}}^{2}+40d_{\textup{c}}-23}}{4(2d_{\textup{c}}^{2}-5d_{\textup{c}}+3)}.
\end{align}
This closed-form equation gives quite close threshold values to the exact ones for various $d_{\textup{c}}$ at $d_{\textup{v}}=3$. The same can be done for $d_{\textup{v}}=4$ as well.

\subsubsection{UDP Setup - Doping}
Observe that the doping equation (\ref{doping_de}) is the same as the uniform equation (\ref{gallager_a}), except the power $d_{\textup{c}}-1-x$ in the former that is $d_{\textup{c}}-1$ in the latter. Similarly, we derive a closed-form approximate equation of the threshold via approximating (\ref{poly_unf}) that is based on (\ref{doping_de}) by taking only $3$ binomial expansion terms to represent $(1-2y)^k$, where $k={d_{\textup{c}}-1-x}$ and $y=p_{-1}^{(0)}$, and then solve the quadratic equation resulting from setting the approximate polynomial to $0$. The smallest solution of (\ref{gal_closed}) with $k = d_{\textup{c}}-1-x$ gives an approximate value of the highest ${p}_{-1}^{(0)}$ for the regular nodes in the doping setup in specific cases. Therefore, $p_{\textup{Doping}}^{\textup{a}} = {p}_{-1}^{(0)}R$, where $R$ is the rate, gives the approximate threshold in the doping setup via averaging on all nodes.

Consider the case of $d_{\textup{v}}=3$ (in which, Gallager A becomes equivalent to Gallager B) and plug $k = d_{\textup{c}}-1-x$ in (\ref{gal_closed}), 
\begin{align}
0 &= (d_{\textup{c}}-4)(4d_{\textup{c}}-18)(p_{-1}^{(0)})^{2} - (d_{\textup{c}}-4)(d_{\textup{c}}-2)p_{-1}^{(0)} +1, \nonumber
\end{align}
\begin{align}
p_{-1}^{(0)} &=  \frac{d_{\textup{c}}^{2}-6d_{\textup{c}}+8 \pm \sqrt{d_{\textup{c}}^{4}-12d_{\textup{c}}^{3}+36d_{\textup{c}}^{2}+40d_{\textup{c}}-224}}{4(2d_{\textup{c}}^{2}-17d_{\textup{c}}+36)}.
\end{align}
After averaging on all nodes, the resulting $p_{\textup{Doping}}^{\textup{a}}=p_{-1}^{(0)}R$ values are quite close to the exact threshold values for various $d_{\textup{c}}$ at $d_{\textup{v}}=3$. The same can be done for $d_{\textup{v}}=4$ as well.

Unlike \cite{exact}, we derive closed-form threshold expressions as a function of code parameters. The accuracy of our approximations ranges between $11-24\%$ as illustrated in Table~\ref{table_gall}. In the table, $p_{\textup{UNF}}$ ($p_{\textup{Doping}}$) is the actual threshold value obtained numerically using iterative DE equations (discussed next). $p_{\textup{UNF}}^{\textup{a}}$ ($p_{\textup{Doping}}^{\textup{a}}$) is the approximate threshold value calculated using the closed form equations we have derived via (\ref{gal_closed}).

\subsection*{UDP-Doping Threshold Gains}
We use DE equations to obtain the threshold numerically. For the doping (resp., uniform) setup, we use equation~(\ref{doping_de}) (resp., (\ref{gallager_a})) and track the evolution of probabilities until the probability of error, ${p}_{-1}^{(l)}$ converges to zero. We search for the highest $p_{-1}^{(0)}$, denoted by ${p}_{-1}^{*(0)}$, that satisfies the convergence constraints. Once again, the threshold for the doping setup is calculated by $p_{\textup{Doping}} = {p}_{-1}^{*(0)}R$. Percentage threshold gain is calculated by $\frac{p_{\textup{UDP/Doping}} - p_{\textup{UNF}}}{ p_{\textup{UNF}}} \times 100\%$.

We illustrate the threshold gains we get by applying our doping idea to various codes on BSC with Gallager A decoder in Table~\ref{table_gall}. As the rate increases, i.e, as the channel crossover probability gets lower, gains decrease as expected. Up to $29.4\%$ gain can be achieved at a rate of $0.80$ by having perfectly reliable parity bits. As we move on to decoders with more quantization, we will observe higher threshold gains across various rate codes.


\section{BSC - 3-level-Alphabet Decoder}\label{sec_bsc_3}
We investigate UDP gains on BSC with a $3$-level decoder. In \cite{de}, authors show that a simple $3$-level decoder (quantized decoder) achieves a close threshold to that of BP decoder (full soft-decision decoder). In this section, our goal is to show that our UDP idea achieves significant gains even with a simple quantized decoder with errors and erasures, that this idea does not require a fully soft-decision decoder, and that the jump in gains from $2$-levels (Section \ref{sec_bsc_b}) to $3$-levels is remarkable. We derive the DE equations for UDP setup followed by iterative approximations, introduce the gains, show consistent gains via decoder simulations, and demonstrate the decoder speed of convergence gains of this UDP idea.

\subsection*{DE Equations for UDP}
Let $\overline{p}_{k}^{(l)}$ ($p_{k}^{(l)}$), for $k  \in \{-1,0,1\}$, denote the probability that the message sent from a reliable (regular) VN to a CN at iteration $l$ equals $k$. Similarly, $\overline{q}_{k}^{(l)}$ ($q_{k}^{(l)}$) denotes the probability that the message sent from a CN to a reliable (regular) VN equals $k$.
\begin{align}\label{eqn_q_bsc_uep}
\widetilde{q}_{0}^{(l)} &= 1-(1 - \overline{p}_{0}^{(l-1)})^{\alpha}(1 - p_{0}^{(l-1)})^{\beta}  \nonumber \\
\widetilde{q}_{1}^{(l)} &= \frac{1}{2} \biggr[(\overline{p}_{1}^{(l-1)} + \overline{p}_{-1}^{(l-1)})^{\alpha}(p_{1}^{(l-1)} + p_{-1}^{(l-1)})^{\beta} \nonumber && \\
&\hspace{+1.0em} +  (\overline{p}_{1}^{(l-1)} - \overline{p}_{-1}^{(l-1)})^{\alpha}(p_{1}^{(l-1)} - p_{-1}^{(l-1)})^{\beta}\biggr] \nonumber \\
\widetilde{q}_{-1}^{(l)} &= \frac{1}{2} \biggr[(\overline{p}_{1}^{(l-1)} + \overline{p}_{-1}^{(l-1)})^{\alpha}(p_{1}^{(l-1)} + p_{-1}^{(l-1)})^{\beta} \nonumber && \\
&\hspace{+1.0em} -  (\overline{p}_{1}^{(l-1)} - \overline{p}_{-1}^{(l-1)})^{\alpha}(p_{1}^{(l-1)} - p_{-1}^{(l-1)})^{\beta}\biggr], 
\end{align}
where for $i  \in \{-1,0,1\}$
\begin{align}\label{eqn_gammai}
(\alpha, \beta) = \left\{\begin{matrix}(x,d_{\textup{c}}-1-x), \textup{ } &\widetilde{q}_{i}^{(l)} = q_{i}^{(l)},
\\ (x-1,d_{\textup{c}}-x), \textup{ } &\widetilde{q}_{i}^{(l)} = \overline{q}_{i}^{(l)}.
\end{matrix}\right.
\end{align}
\begin{align} \label{eqn_p_bsc_uep}
\widetilde{p}_{k'}^{(l)} &= \widetilde{p}_{0}^{(0)}\hspace{-1.2em}\sum_{(i,j):i-j \bigtriangleup 0}\hspace{-1em}\Phi  \hspace{0.4em}+ \hspace{0.4em}\widetilde{p}_{1}^{(0)} \hspace{-1.5em}\sum_{(i,j):i-j \bigtriangleup -w_{l}}\hspace{-1.5em}\Phi \hspace{0.4em}+ \hspace{0.4em}\widetilde{p}_{-1}^{(0)} \hspace{-1.2em}\sum_{(i,j):i-j \bigtriangleup w_{l}}\hspace{-1.2em}\Phi \nonumber \\
\widetilde{p}_{-1}^{(l)} &= 1-\widetilde{p}_{0}^{(l)}-\widetilde{p}_{1}^{(l)},
\end{align}
where $k' \in \{0,1\}$, $\Phi \hspace{-0.3em} := \hspace{-0.3em}\binom{d_{\textup{v}}-1}{i,j,d_{\textup{v}}\hspace{-0.1em}-\hspace{-0.1em}1\hspace{-0.1em}-\hspace{-0.1em}i\hspace{-0.1em}-\hspace{-0.1em}j} (\widetilde{q}_{1}^{(l)})^{i}(\widetilde{q}_{-1}^{(l)})^{j}(\widetilde{q}_{0}^{(l)})^{d_{\textup{v}}\hspace{-0.1em}-\hspace{-0.1em}1\hspace{-0.1em}-\hspace{-0.1em}i\hspace{-0.1em}-\hspace{-0.1em}j}$,  and the operator $\bigtriangleup$ is equality ($=$) if $k'=0$ and is greater than ($>$) if $k'=1$. In (\ref{eqn_p_bsc_uep}), for reliable VNs, $\widetilde{p}_{i}^{(l)} = \overline{p}_{i}^{(l)}$ and  $\widetilde{q}_{i}^{(l)} = \overline{q}_{i}^{(l)}$; for regular VNs, $\widetilde{p}_{i}^{(l)} = p_{i}^{(l)}$ and  $\widetilde{q}_{i}^{(l)} = q_{i}^{(l)}$ where $i \in \{-1,0,1\}$.


\subsection*{Approximations}
The evolution of $\widetilde{p}_{-1}^{(l)}$ can be further approximated as a function of $\widetilde{p}_{-1}^{(0)}$, $\widetilde{p}_{-1}^{(l-1)}$, $\widetilde{p}_{1}^{(l-1)}$, and code parameters ($d_{\textup{v}}, d_{\textup{c}}$). We can substitute (\ref{eqn_q_bsc_uep}) in (\ref{eqn_p_bsc_uep}) and obtain,
\begin{align}
\widetilde{p}_{-1}^{(l)} &= 1- \widetilde{p}_{1}^{(0)}\biggr[\frac{1}{2}
\Omega^{2} - \frac{1}{2}\Lambda^{2} - \Omega - \Lambda + \Omega\Lambda + 1 \biggr] \nonumber \\
&\hspace{+1.0em}+ \frac{3}{4}\Omega^{2} - \frac{1}{4}\Lambda^{2} - \Omega - \Lambda + \frac{1}{2}\Omega\Lambda,
\end{align}
where 
\begin{align}
\Omega &= (\overline{p}_{1}^{(l-1)} + \overline{p}_{-1}^{(l-1)})^{\alpha}({p}_{1}^{(l-1)} + {p}_{-1}^{(l-1)})^{\beta}, \nonumber \\
\Lambda &= (\overline{p}_{1}^{(l-1)} - \overline{p}_{-1}^{(l-1)})^{\alpha}({p}_{1}^{(l-1)} - {p}_{-1}^{(l-1)})^{\beta}. 
\end{align}
Here, $\widetilde{p}_{1}^{(l)}$ converges to $1$, while $\widetilde{p}_{-1}^{(l)}$ converges to $0$ in successful decoding. Thus, we apply the following approximations (by taking only $2$ binomial expansion terms),
\begin{align}
(\widetilde{p}_{1}^{(l)} \pm \widetilde{p}_{-1}^{(l)})^{\alpha} &\approx (\widetilde{p}_{1}^{(l)})^{\alpha} \pm \alpha(\widetilde{p}_{1}^{(l)})^{\alpha-1}\widetilde{p}_{-1}^{(l)}, \nonumber \\
\bigr((\widetilde{p}_{1}^{(l)})^{2} - (\widetilde{p}_{-1}^{(l)})^{2}\bigr)^{\alpha} &\approx (\widetilde{p}_{1}^{(l)})^{2\alpha} - \alpha(\widetilde{p}_{1}^{(l)})^{2(\alpha-1)}(\widetilde{p}_{-1}^{(l)})^{2}.
\end{align}

Thus, over a BSC with $3$-level decoder, the evolution of $\widetilde{p}_{-1}^{(l)}$ for a regular LDPC code where the UDP idea is applied can be approximated using (\ref{bsc_approx_uep1}) and (\ref{bsc_approx_uep2}) (for simplicity, we consider the case of $w_{l}=1$, $d_{\textup{v}}=3$). In Table \ref{table1_3l}, the approximate threshold values obtained, $p_{\textup{UDP}}^{\textup{a}}$, are illustrated. These approximations offer accuracy within $16\%$.


\begin{figure*}

\begin{align} 
\widetilde{p}_{-1}^{(l)} &\approx \widetilde{p}_{-1}^{(0)}\biggr[ A - 2({\overline{p}_{1}^{(l-1)}})^{\alpha}({p_{1}^{(l-1)}})^{\beta} - 2\alpha\beta({\overline{p}_{1}^{(l-1)}})^{\alpha-1}({p_{1}^{(l-1)}})^{\beta-1}\overline{p}_{-1}^{(l-1)}p_{-1}^{(l-1)} + B \biggr]  + C  + D \label{bsc_approx_uep1} \\
\widetilde{p}_{1}^{(l)} &\approx (1\hspace{-0.1em}-\hspace{-0.1em}\widetilde{p}_{-1}^{(0)})\biggr[ A - 2({\overline{p}_{1}^{(l-1)}})^{\alpha}\beta ({{p}_{1}^{(l-1)}})^{\beta-1}p_{-1}^{(l-1)} - 2\alpha({\overline{p}_{1}^{(l-1)}})^{\alpha-1}({\overline{p}_{-1}^{(l-1)}}) ({{p}_{1}^{(l-1)}})^{\beta} \hspace{-0.1em}-\hspace{-0.1em} B \biggr] + ({\overline{p}_{1}^{(l-1)}})^{2\alpha}({{p}_{1}^{(l-1)}})^{2\beta} \hspace{-0.1em}+\hspace{-0.1em} C \hspace{-0.1em}- \hspace{-0.1em}D ,\label{bsc_approx_uep2}
\end{align}
where,
\begin{align}
A &= 1 + 2({\overline{p}_{1}^{(l-1)}})^{2\alpha}\beta({p_{1}^{(l-1)}})^{2\beta-1}{p_{-1}^{(l-1)}} + 2\alpha({\overline{p}_{1}^{(l-1)}})^{2\alpha-1} {\overline{p}_{-1}^{(l-1)}}({p_{1}^{(l-1)}})^{2\beta}, \hspace{15em} \nonumber \\
B &= \bigr(({\overline{p}_{1}^{(l-1)}})^{2\alpha} - \alpha({\overline{p}_{1}^{(l-1)}})^{2(\alpha-1)}({\overline{p}_{-1}^{(l-1)}})^{2} \bigr)\bigr( ({p_{1}^{(l-1)}})^{2\beta} - \beta({{p}_{1}^{(l-1)}})^{2(\beta-1)}(p_{-1}^{(l-1)})^{2}\bigr), \nonumber \\ 
C &= 2\alpha\beta({\overline{p}_{1}^{(l-1)}})^{2\alpha-1}({p_{1}^{(l-1)}})^{2\beta-1}\overline{p}_{-1}^{(l-1)}p_{-1}^{(l-1)}, \hspace{1em} \textup{and } \hspace{1em} D = \frac{1}{2}({\overline{p}_{1}^{(l-1)}})^{2\alpha} ({p_{1}^{(l-1)}})^{2\beta} - \frac{1}{2}B. \nonumber
\end{align}
\rule{18cm}{0.4pt}
\end{figure*}


\subsection*{UDP Threshold Gains}
We use DE equations to obtain the threshold of a regular $(d_{\textup{v}},d_{\textup{c}})$ LDPC code  numerically. For UDP setup, we use (\ref{eqn_q_bsc_uep}) and (\ref{eqn_p_bsc_uep}), and track the evolution of probabilities until the probability of error $\widetilde{p}_{-1}^{(l)}$ and probability of erasure $\widetilde{p}_{0}^{(l)}$ converge to zero for both reliable and regular VNs for each $(p_{-1}^{(0)}, \overline{p}_{-1}^{(0)})$ pair. We search for the best pair with the highest average that satisfies the convergence constraints, $(p_{-1}^{*(0)}, \overline{p}_{-1}^{*(0)})$. For example in $(3,12)$ code, $(p_{-1}^{*(0)}, \overline{p}_{-1}^{*(0)}) = (0.033,0.0009)$ and in $(5,10)$ code, $(p_{-1}^{*(0)}, \overline{p}_{-1}^{*(0)}) = (0.147,0.0003)$. Then, the threshold for UDP setup is calculated as follows:
\begin{align}
    p_{\textup{UDP}} &= \overline{p}_{-1}^{*(0)}(1-R) + p_{-1}^{*(0)}R. \label{eqn_th_uep}
\end{align}
Similarly, for uniform setup, we use (\ref{eqn_q_bsc_unif}) and (\ref{eqn_p_bsc_unif}) to obtain the code threshold $p_{\textup{UNF}} = p_{-1}^{*(0)}$ when error and erasure probabilities converge to zero.

\begin{table}
\caption{Threshold Gains of UDP for Various $(d_{\textup{v}},d_{\textup{c}})$ LDPC Codes On BSC with $3$-Level Decoder}
\centering
\begin{tabular}{|c|c|c|c|c|c|c|}
\hline
$d_{\textup{v}}$ & $d_{\textup{c}}$ & Rate & $p_{\textup{UNF}}$ & $p_{\textup{UDP}}$ & $p^{\textup{a}}_{\textup{UDP}}$ & \textbf{Gain}   \\
\specialrule{.1em}{.05em}{.05em} 
$3$ & $8$ & $0.63$ & $0.0421$ & $0.0582$ & $0.0492$ & $38.2\%$ \\
\hline
$3$ & $10$ & $0.70$ & $0.0283$ & $0.0366$ & $0.0338$ & $29.3\%$ \\
\hline
$3$ & $12$ & $0.75$ & $0.0205$ & $0.0250$ & $0.0248$ & $22.0\%$ \\
\hline
$3$ & $15$ & $0.80$ & $0.0139$ & $0.0161$ & $0.0168$ & $15.8\%$ \\
\hline
$3$ & $27$ & $0.89$ & $0.0052$ & $0.0055$ & $0.0057$ & $5.8\%$ \\
\specialrule{.1em}{.05em}{.05em} 
$4$ & $10$ & $0.60$ & $0.0439$ & $0.0529$ & -- & $20.5\%$ \\
\hline
$4$ & $12$ & $0.67$ & $0.0345$ & $0.0389$ & -- & $12.8\%$ \\
\hline
$4$ & $16$ & $0.75$ & $0.0238$ & $0.0256$ & -- & $7.6\%$ \\
\specialrule{.1em}{.05em}{.05em} 
$5$ & $10$ & $0.50$ & $0.0552$ & $0.0737$ & -- & $33.5\%$ \\
\hline
$5$ & $15$ & $0.67$ & $0.0311$ & $0.0363$ & -- & $16.7\%$ \\
\hline
$5$ & $25$ & $0.80$ & $0.0159$ & $0.0170$ & -- & $6.9\%$ \\
\hline
\end{tabular}
\label{table1_3l}
\end{table}

We illustrate the threshold gains we get by applying our UDP idea to various codes on BSC with $3$-level decoder in Table~\ref{table1_3l}. As the rate increases, i.e, as the system gets better (lower crossover error probabilities), gains decrease as expected. The results demonstrate that UDP offers a remarkable gain of around $38.2\%$ for a moderate rate of around $0.63$, which is suitable for wireless communication. This threshold gain means that with our UDP setup, the exact same code can appropriately operate at average error probabilities that are up to $38.2\%$ higher than the maximum it can operate at under the uniform setup. There are also gains at higher rates suitable for data storage. At data storage rates of around $0.75$ and $0.80$, with $(3,12)$ and $(3,15)$ codes, significant gains of $22\%$ and $15.8\%$ are achieved, respectively. These results are a further demonstration that UDP via more reliable parity bits is a promising idea that achieves significant threshold gains.

\begin{figure}
\vspace{-1.0em}
\centering
\includegraphics[trim={1.2in 0.5in 1.2in 0.5in},width=3.4in]{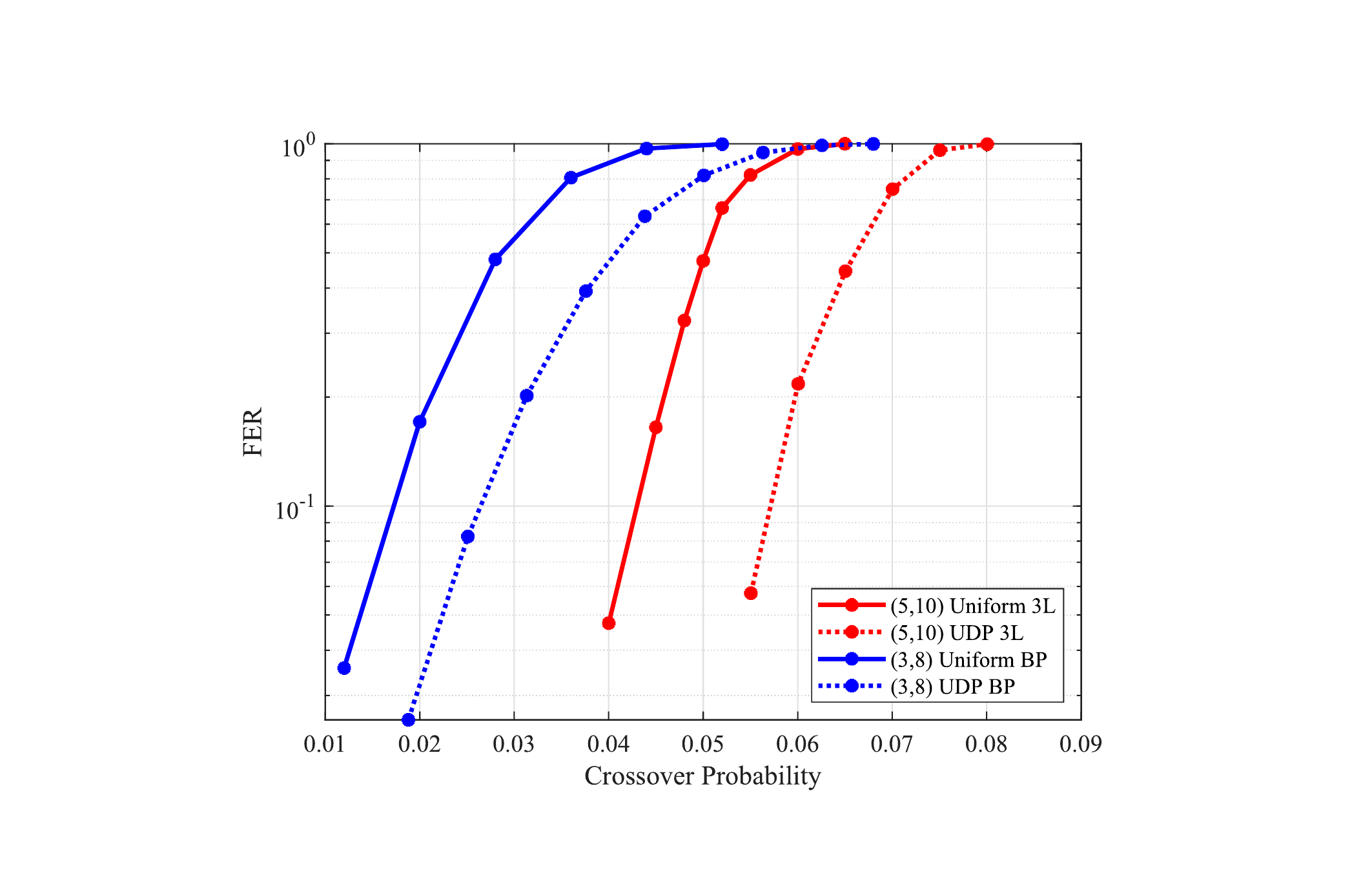}
\caption{FER versus crossover probability, finite-length simulations of $(5,10)$ code with on $3$-Level Decoder (red) with block length $3020$ and $(3,8)$ code on BP Decoder (blue) with block length $3632$.}
\label{fig_1}
\end{figure}

We verified the gains illustrated in Table~\ref{table1_3l} by implementing the BSC 3-level decoder and simulating finite-length regular LDPC codes. We find the frame error rate (FER) at different crossover probabilities of regular LDPC codes with both uniform and UDP setups applied via simulations. The x-axis in the figure is the crossover probability in the uniform case, while it is the average crossover probability from (\ref{eqn_th_uep}) in the UDP case. Figure~\ref{fig_1} (red curves) shows around $30.4\%$ gain obtained at around FER $0.82$ (FER at the threshold value reported in Table~\ref{table1_3l} for uniform setup) of a $(5,10)$ LDPC code when the UDP idea is applied. This result is significant because it demonstrates that the threshold gains obtained using infinite asymptotic analysis via DE are consistent with the finite-length decoder simulations. We also observe that LDPC decoder experiences a version of the channel with a higher effective SNR.

\subsection*{UDP Speed of Convergence Gains}
Next, we illustrate the speed of convergence gains of our UDP idea on the $3$-level decoder. At the same error probability (average error probability for the UDP setup), the average number of iterations needed for the decoder to converge is calculated for both setups. Codewords for which the decoder does not converge at all are not included on the average. The results are illustrated in Table~\ref{table_s_3l}. Codes with block lengths in the range $3020-14292$ are simulated.

\begin{table}
\caption{Speed of Convergence Gains of UDP for Various $(d_{\textup{v}},d_{\textup{c}})$ LDPC Codes On BSC with $3$-Level Decoder}
\centering

\begin{tabular}{|c|c|c|c|c|c|}
\hline
\multirow{2}{*}{Code} & \multirow{2}{*}{$p_{-1}^{(0)}$} & \multirow{2}{*}{\shortstack{FER\\ (UNF)}} & \multirow{2}{*}{\shortstack{FER\\ (UDP)}} & \multirow{2}{*}{\shortstack{Iteration\\ Count (UNF)}} & \multirow{2}{*}{\shortstack{Iteration\\ Count (UDP)}}  \\
& & & & & \\
\specialrule{.1em}{.05em}{.05em} 
\multirow{2}{*}{$(5,10)$} & $0.048$ & $0.325$ & $0.015$ & $18.45$ & $12.69$  \\
\cline{2-6}
 & $0.052$ & $0.665$ & $0.026$ & $21.13$ & $15.13$  \\
\specialrule{.1em}{.05em}{.05em}  
\multirow{2}{*}{$(4,8)$} & $0.047$ & $0.512$ & $0.007$ & $17.68$ & $10.22$  \\
\cline{2-6}
 & $0.049$ & $0.630$ & $0.009$ & $18.38$ & $10.61$ \\
 \specialrule{.1em}{.05em}{.05em}  
\multirow{2}{*}{$(4,10)$} 
 & $0.033$ & $0.428$ & $0.106$ & $25.05$ & $23.13$  \\
\cline{2-6}
 & $0.035$ & $0.620$ & $0.180$ & $27.05$ & $24.36$  \\
\specialrule{.1em}{.05em}{.05em} 
\multirow{2}{*}{$(3,18)$} & $0.006$ & $0.326$ & $0.210$ & $14.92$ & $12.26$  \\
\cline{2-6}
 & $0.007$ & $0.468$ & $0.386$ & $17.81$ & $16.88$ \\
\hline
\end{tabular}
\label{table_s_3l}
\end{table}

\subsection*{Extreme Case - Doping, Threshold Gains}
Throughout this section, we have explored the performance gains of the UDP idea where information and parity bits have different reliabilities. An extreme case of this idea is when the parity bits are perfectly reliable (known to the decoder), i.e., $\overline{p}_{-1}^{(l)} = \overline{p}_{0}^{(l)} = 0$ and $\overline{p}_{1}^{(l)} = 1$, for all iterations. In Table~\ref{table_doping_3l}, threshold gains obtained by UDP and doping versus uniform are compared. As expected, when parity bits perfectly reliable, higher threshold gains can be achieved.

\begin{table}
\caption{Threshold Gains of UDP and Doping for Various $(d_{\textup{v}},d_{\textup{c}})$ LDPC Codes On BSC with $3$-Level Decoder}
\centering
\begin{tabular}{|c|c|c|c|c|c|c|}
\hline
$d_{\textup{v}}$ & $d_{\textup{c}}$ &  $p_{\textup{UNF}}$ & $p_{\textup{UDP}}$ & $p_{\textup{DOPING}}$ & $\textup{Gain}_{\textup{UDP}}$ & $\textup{Gain}_{\textup{Doping}}$ \\
\specialrule{.1em}{.05em}{.05em} 
$3$ & $8$ & $0.0421$ & $0.0582$ & $0.0598$ & $38.2\%$ & $42.0\%$ \\
\hline
$3$ & $9$ & $0.0341$ & $0.0455$ & $0.0466$ & $33.4\%$ & $36.7\%$ \\
\hline
$3$ & $15$ & $0.0139$ & $0.0161$ & $0.0164$ & $15.8\%$ & $18.0\%$ \\
\hline
$3$ & $27$ & $0.0052$ & $0.0055$ & $0.0056$ & $5.8\%$ & $7.7\%$ \\
\specialrule{.1em}{.05em}{.05em} 
$4$ & $10$ & $0.0439$ & $0.0529$ & $0.0542$ & $20.5\%$ & $23.5\%$ \\
\hline
$4$ & $16$ & $0.0238$ & $0.0256$ & $0.0259$ & $7.6\%$ & $8.8\%$ \\
\specialrule{.1em}{.05em}{.05em} 
$5$ & $15$ & $0.0311$ & $0.0363$ & $0.0368$ & $16.7\%$ & $18.3\%$ \\
\hline
\end{tabular}
\label{table_doping_3l}
\end{table}

\section{BSC - BP Decoder}\label{sec_bsc_bp}
We now investigate UDP gains on BSC with a BP decoder. In order to highlight the effectiveness and validity of our UDP idea, we show that UDP achieves significant gains under the widely used BP decoder. In this section, we introduce the DE equations for UDP setup, illustrate the threshold gains, demonstrate the FER gains via BP decoder simulations, and show the decoder speed of convergence gains of UDP.

\subsection*{DE Equations for UDP}

The message sent from a VN to a CN is the sum of the channel message and $(d_{\textup{v}}-1)$ messages coming from other adjacent CNs. In UDP setup, reliable and regular VNs have different channel contributions, and the message sent from a CN to a VN also differs depending on whether the VN is reliable or regular. The independence assumption holds across the transferred messages among nodes. Probability density of the message sent from a VN to a CN is calculated by convolution as in uniform setup.

When the CN-to-VN message is considered in our UDP setup, the messages going to the CN from edge $t \in [d_{\textup{c}}-1]$ have different probability densities depending on whether the message is coming from $x$ reliable VNs, denoted by $\overline{p}_{i}^{t}$, or from $d_{\textup{c}}-x$ regular VNs, denoted by ${p}_{i}^{t}$ ($i\in \{0,1\}$). The outgoing message from a CN along an edge is the LLR $\log\frac{\overline{p}'_{0}}{\overline{p}'_{1}}$ ($\log\frac{p'_{0}}{p'_{1}}$) if it is going to a reliable (regular) VN. The new update law is specified by:
\vspace{-0.1em}\begin{align} 
\widetilde{p}'_{0} - \widetilde{p}'_{1}&= \hspace{-0.2em}\prod_{t=1}^{d_{\textup{c}}-1} \big(\overline{p}_{0}^{t} - \overline{p}_{1}^{t} \big)^{\alpha}\big(p_{0}^{t} - p_{1}^{t} \big)^{d_{\textup{c}}-1-\alpha}, \end{align}
where $\alpha = x-1$ $(\alpha= x)$ if $\widetilde{p}'_{i} = \overline{p}'_{i}$ ($\widetilde{p}'_{i} = {p}'_{i}$). 

In order to study the evolution of densities at the CN side, in UDP, the representation in (\ref{representation}) becomes:
\begin{align} 
(\textup{lgsgn}(\overline{p}_{0}-\overline{p}_{1}), -\log|\overline{p}_{0}-\overline{p}_{1}|), \label{rep_rel}\\
(\textup{lgsgn}(p_{0}-p_{1}), -\log|p_{0}-p_{1}|). \label{rep_reg}
\end{align} 
We define $\overline{L}=\log\frac{\overline{p}_{0}}{\overline{p}_{1}}$ and $L=\log\frac{p_{0}}{p_{1}}$ for reliable and regular VNs, respectively. Therefore,  $\overline{p}_{0}-\overline{p}_{1} = \frac{e^{\overline{L}}-1}{e^{\overline{L}}+1}$ and $p_{0}-p_{1} = \frac{e^{L}-1}{e^{L}+1}$ can be substituted in (\ref{rep_rel}) and (\ref{rep_reg}), respectively.
Once again, with this representation, the CN-to-VN message can be computed via addition. However, in UDP, the summation will include the reliable representation in (\ref{rep_rel}) as well as the regular representation in  (\ref{rep_reg}) with appropriate bounds specified by $\alpha$. The rest of the steps will be the same as in the uniform case.


\subsection*{UDP Threshold Gains}


We implemented DE software for the BP decoder to find thresholds of both uniform and UDP setups. Guided by \cite{de}, we have followed the methodology described above and in Section~\ref{de_bp_uniform} to obtain the code threshold for UDP and uniform setups, respectively. For UDP, we search for the best $(p_{-1}^{(0)}, \overline{p}_{-1}^{(0)})$ combination with the highest average, $(p_{-1}^{*(0)}, \overline{p}_{-1}^{*(0)})$, that satisfies the convergence constraints using Newton's method.

\begin{table}
\caption{Threshold Gains of UDP for Various $(d_{\textup{v}},d_{\textup{c}})$ LDPC Codes On BSC with BP Decoder}
\centering

\begin{tabular}{|c|c|c|c|c|c|c|c|}
\hline
$d_{\textup{v}}$ & $d_{\textup{c}}$ & Rate & $p_{\textup{UNF}}$ & $p_{\textup{UDP}}$ & \textbf{Gain}   \\ 
\specialrule{.1em}{.05em}{.05em} 
$3$ & $7$ & $0.57$ & $0.0625$ & $0.0892$  & $42.7\%$ \\ 
\hline
$3$ & $8$ & $0.63$ & $0.0505$ & $0.0673$ & $33.3\%$ \\ 
\hline
$3$ & $10$ & $0.70$ & $0.0357$ & $0.0435$ &  $21.8\%$ \\ 
\hline
$3$ & $12$ & $0.75$ & $0.0270$ & $0.0314$ &  $16.3\%$ \\ 
\hline
$3$ & $15$ & $0.80$ & $0.0193$ & $0.0215$ &  $11.4\%$ \\ 
\hline
$3$ & $18$ & $0.83$ & $0.0147$ & $0.0160$ & $8.8\%$ \\ 
\specialrule{.1em}{.05em}{.05em}
$4$ & $8$ & $0.50$ & $0.0705$ & $0.0934$ &  $32.5\%$ \\ 
\hline
$4$ & $10$ & $0.60$ & $0.0517$ & $0.0624$ &  $20.7\%$ \\ 
\hline
$4$ & $12$ & $0.67$ & $0.0402$ & $0.0469$ &  $16.7\%$ \\ 
\hline
$4$ & $20$ & $0.80$ & $0.0200$ & $0.0216$ &  $8.0\%$ \\ 
\specialrule{.1em}{.05em}{.05em} 
$5$ & $15$ & $0.67$ & $0.0358$ & $0.0401$ &  $12.0\%$ \\ 
\hline
$5$ & $20$ & $0.75$ & $0.0247$ & $0.0268$ &  $8.5\%$ \\ 
\hline
$5$ & $25$ & $0.80$ & $0.0186$ & $0.0198$ &  $6.5\%$ \\ 
\hline
\end{tabular}
\label{table1_bp}
\end{table}


We illustrate the UDP threshold gains we get on BSC with BP decoder in Table~\ref{table1_bp}\footnote{In our BP DE software, minor discrepancies from reported literature values may be observed for uniform threshold values. This is due to FFT/IFFT implementations in BP software. Regardless, the same discrepancies apply to UDP values. Therefore, relative gains remain the same.}. In a way similar to Table~\ref{table1_3l}, at moderate and lower rates corresponding to higher crossover probabilities, we obtain higher gains. At a rate of around $0.57$ with $(3,7)$ code, which is suitable for wireless communication, $42.7\%$ gain is obtained with higher protection of parity bits. At data storage rates of around $0.75$ with $(3,12)$ code and $0.80$ with $(3,15)$ code, significant gains of $16.3\%$ and $11.4\%$ are achieved, respectively. At a moderate rate of around $0.70$ with $(3,10)$ code, almost $22\%$ gain is obtained. This gain means that with our UDP setup, the exact same code can appropriately operate at average crossover probabilities that are up to $22\%$ higher than the maximum it can operate at under uniform setup. In data storage terms, as the device deteriorates, lower rate codes are typically used to combat higher error probabilities and increase lifetime. Whereas with UDP idea, lifetime gains can be achieved with no (or limited) rate loss.

We verified the gains illustrated in Table~\ref{table1_bp} by simulating finite-length regular LDPC codes on BSC with BP decoder. Figure~\ref{fig_1} (blue curves) shows around $27.7\%$ gain obtained at around FER $0.99$ (FER at the threshold value reported in Table~\ref{table1_bp} for uniform setup) of $(3,8)$ LDPC code when the UDP idea is applied compared with the uniform setup. Thus, gains obtained from asymptotic analysis are consistent with the finite-length decoder simulations here as well.

\subsection*{UDP Speed of Convergence Gains}
Finally, we demonstrate the speed of convergence gains of our UDP idea on the BP decoder in Table~\ref{table_s_bp}. Codes with block lengths in the range $10576-30548$ are simulated. At low rates (the high gain region) better speed of convergence results are obtained, which is consistent with the threshold gains. These speed of convergence gains are translated into notable power savings in practical systems.

\begin{table}
\caption{Speed of Convergence Gains of UDP for Various $(d_{\textup{v}},d_{\textup{c}})$ LDPC Codes On BSC with BP Decoder}
\centering

\begin{tabular}{|c|c|c|c|c|c|}
\hline
\multirow{2}{*}{Code} & \multirow{2}{*}{$p_{-1}^{(0)}$} & \multirow{2}{*}{\shortstack{FER\\ (UNF)}} & \multirow{2}{*}{\shortstack{FER\\ (UDP)}} & \multirow{2}{*}{\shortstack{Iteration\\ Count (UNF)}} & \multirow{2}{*}{\shortstack{Iteration\\ Count (UDP)}}  \\
& & & & & \\
\specialrule{.1em}{.05em}{.05em} 
\multirow{3}{*}{$(5,10)$} & $0.056$ & $0.455$ & $0.067$ & $20.69$ & $9.91$  \\
\cline{2-6}
 & $0.058$ & $0.565$ & $0.081$ & $23.66$ & $10.89$ \\
 \cline{2-6}
 & $0.062$ & $0.700$ & $0.146$ & $29.40$ & $12.67$ \\
\specialrule{.1em}{.05em}{.05em} 
\multirow{2}{*}{$(4,8)$} & $0.055$ & $0.425$ & $0.010$ & $16.85$ & $6.86$ \\
\cline{2-6}
 & $0.060$ & $0.601$ & $0.018$ & $21.23$ & $7.89$ \\
\specialrule{.1em}{.05em}{.05em} 
\multirow{2}{*}{$(4,14)$} & $0.027$ & $0.324$ & $0.238$ & $23.91$ & $20.67$  \\
\cline{2-6}
 & $0.028$ & $0.413$ & $0.298$ & $26.30$ & $22.02$ \\
\specialrule{.1em}{.05em}{.05em} 
\multirow{2}{*}{$(3,18)$} & $0.008$ & $0.297$ & $0.264$ & $9.57$ & $9.16$  \\
\cline{2-6}
 & $0.010$ & $0.565$ & $0.514$ & $13.79$ & $12.66$ \\
\hline
\end{tabular}
\label{table_s_bp}
\end{table}

\section{Conclusion}\label{sec_conc}
We provided threshold analysis of our UDP idea, characterized by more reliable parity bits, on BSC with finite to infinite alphabet decoders. For Gallager A, $3$-level, and BP decoders, we derived DE equations for UDP and doping setups. We further analyzed the iterative algorithm performance to provide insights into decoder dynamics. In certain cases, we theoretically derived exact thresholds, tight closed-form approximations that are function of code parameters, as well as iterative approximations. Up to $38.2\%$ and $42.7\%$ threshold gains are achieved for $3$-level and BP decoders, respectively. Finite-length decoder simulations confirm the asymptotic gains computed from the DE analysis. UDP offers better decoder performance by achieving a higher speed of convergence at the same noise level. We observed improvement in gains moving from $2$ levels in Gallager A decoding to $3$ levels and BP decoding. With UDP and doping, the LDPC decoder experiences a version of the channel with a higher effective SNR. We suggest that this setup can contribute to significant density/lifetime gains in various data storage systems and transmission gains in wireless communication systems.


\section*{Acknowledgment}\label{sec_ack}
The authors would like to thank Siyi Yang for her assistance in carrying out this research. This work was supported in part by the TUBITAK 2232-B International Fellowship for Early Stage Researchers.


\end{document}